\documentclass{PoS}

\usepackage{graphicx}
\usepackage{setspace}
\usepackage{amsmath}
\usepackage{color}
\usepackage{epstopdf}

\def\be{\begin{equation}}
\def\ee{\end{equation}}
\def\bea{\begin{eqnarray}}
\def\eea{\end{eqnarray}}

\newcommand{\calo}{{\cal O}}

\newcommand{\Tr}{{\rm Tr}\,}

\newcommand{\ax}{a_x}
\newcommand{\axd}{a_x^\dagger}
\newcommand{\bx}{b_x}
\newcommand{\bxd}{b_x^\dagger}
\newcommand{\ay}{a_y}

\newcommand{\by}{b_y}

\title{On the semimetal-insulator transition and Lifshitz transition in
simulations of mono-layer graphene}

\ShortTitle{Semimetal-insulator transition and Lifshitz transition in 
mono-layer graphene}

\author{\speaker{Dominik Smith}
       \\
        Theoriezentrum, Institut f\"ur Kernphysik, TU Darmstadt,
        64289 Darmstadt, Germany\\
        E-mail: \email{smith@theorie.ikp.physik.tu-darmstadt.de} }

\author{Michael K\"orner 
       \\
        Theoriezentrum, Institut f\"ur Kernphysik, TU Darmstadt,
        64289 Darmstadt, Germany\\
        E-mail: \email{koerner@theorie.ikp.physik.tu-darmstadt.de} }

\author{Lorenz von Smekal\\
        Theoriezentrum, Institut f\"ur Kernphysik, TU Darmstadt,
        64289 Darmstadt, Germany\\
        Institut f\"ur Theoretische Physik, Justus-Liebig-Universit\"at,
        35392 Giessen, Germany\\
        E-mail: \email{lorenz.smekal@physik.tu-darmstadt.de}}

\abstract{We report on the status of ongoing Hybrid-Monte-Carlo simulations of the
tight-binding model of mono-layer graphene. We present results concerning the
semimetal-insulator phase transition, whereby two-body interactions are
modeled by a partially screened Coulomb potential which takes into account
screening by electrons in the lower $\sigma$-orbitals. We obtain evidence that
finite-size effects may still be present in the current estimate of 
the critical coupling strength $\alpha_C$, which was previously extracted from simulations 
on lattice-sizes up to $N_x=N_y=18$. 
We also present
preliminary results concerning the Neck-disrupting Lifshitz transition which
occurs at finite Fermion-density in the limit of vanishing two-body interactions.
A sign-problem is circumvented by using a spin-dependent chemical potential in
our simulations. }

\FullConference{The 32nd International Symposium on Lattice Field Theory\\
                 23-28 June, 2014\\
                 Columbia University New York, NY}

\begin{document}

\section{Introduction}

In recent years, much interest has arisen in graphene, a two-dimensional
sheet of carbon atoms arranged on a hexagonal lattice, due to its many unusual
properties which make it an interesting candidate for a number of technological
applications \cite{CastroNeto:2009zzKotov}. From a theoretical perspective, graphene is interesting
because its electronic properties are governed by a massless Dirac equation
in the limit of low energies and because the small Fermi velocity 
$v_F \approx c/300$ leads to strong electromagnetic interactions with an effective fine-structure
constant of $\alpha_{\textrm{eff}}=e^2/(\hbar v_F)\approx 2.2$ \cite{Gusynin:2007ix}. Thus, graphene
provides an example of the physics of strongly coupled relativistic field theory, realized
in a condensed matter system. Furthermore, the coupling-constant in graphene can be directly manipulated by affixing the sheet to a substrate, which leads to di-electric rescaling $\alpha_{\textrm{eff}} \to \alpha_{\textrm{eff}}/\epsilon$.

Early on, the lattice community took an interest in graphene since it seemed obvious
that the well-developed machinery for simulating field theories should be applicable. Several works were
published which focused on electronic properties
in the low-energy limit (see e.g. Refs. \cite{Drut:2008rgArmour:2009vj}).
Later it was shown how the full microscopic theory with hexagonal
symmetry can be formulated as a lattice theory \cite{Brower:2012zd}. This
was henceforth adopted as the de-facto standard and a number of works
were published presenting simulations based on this derivation
\cite{Buividovich:2012nx,Ulybyshev:2013swa,Smith:2013pxa,Smith:2014tha}.

In this work we report on our ongoing efforts to simulate the electronic
properties of graphene via Hybrid-Monte-Carlo. We implement a fully hexagonal
lattice theory based on the derivations presented in Ref. \cite{Brower:2012zd}.

The aim of this paper is two-fold: After summarizing the methodology,
we first present a follow-up discussion of our investigation
of the semimetal-insulator phase-transition which was published in
Ref. \cite{Smith:2014tha}. Therein simulations of the interacting
tight-binding theory of graphene were presented in which electronic
two-body interactions were modeled by a partially screened Coulomb
potential, which accounts for screening by electrons in the lower
$\sigma$-orbits at short distances and crosses over smoothly
to an unscreened Coulomb potential at long distances. 
We contrasted
these results to a similar study presented in Ref. \cite{Ulybyshev:2013swa},
which instead assumed that at long distances the potential is screened
by a constant di-electric screening factor. We found that for a lattice
of dimensions $N_x=N_y=18$ the exact form of the long-range tail doesn't seem
to affect the precise location of the critical coupling strength $\alpha_C$ for
gap-formation much. In this work we present results obtained on $N_x=N_y=36$
which may indicate that for larger systems the transition is shifted to smaller
$\alpha_C$. This would mean that previous simulations are affected by
finite-size effects and that an extrapolation to infinite surface area
is in order to conclusively decide whether the long-range part of the potential
is of relevance. 

Second we present preliminary results concerning the
Neck-disrupting Lifshitz transition, which is known to occur in the
limit of vanishing two-body interactions at finite Fermion-density
and which is characterized by a change of the topology of the Fermi 
surface and a logarithmic divergence of the density of states \cite{Dietz:2013sga}.
To avoid a Fermion sign-problem we implement the ``isospin'' chemical
potential, the sign of which differs for the two spin-orientations of
electrons. The results presented here concern the non-interacting limit only.
Our long-term goal is to investigate how a small interaction term affects
the Lifshitz transition.

\section{The setup}

The setup of our simulations has been described in great
detail in Ref. \cite{Smith:2014tha}, where a step-by-step derivation
of every component of the Hybrid-Monte-Carlo algorithm for the interacting 
tight-binding theory based on the concepts worked out in Ref. \cite{Brower:2012zd}
is presented. We will only summarize the basic principles here. Our goal is to
simulate the thermodynamics of the interacting tight-binding theory of graphene
\begin{equation}
H = \sum_{\langle x,y \rangle}(-\kappa)(\axd \ay - \bxd \by + \textrm{h.c.}) 
+ \sum_{x,y}\, q_x V_{xy} q_y +  \sum_{x}m_s (\axd \ax + \bxd \bx)~,
\end{equation}
where $\ax,\axd,\bx,\bxd$ are ladder operators for particles with spin $+1/2$ and
anti-particles (``holes'') with spin $-1/2$ respectively. 
The sums run over all pairs of nearest neighbors, pairs of coordinates and all coordinates of the hexagonal lattice
respectively. $q_x = \axd \ax - \bxd \bx$ is the charge operator, $\kappa\approx 2.7$ 
the hopping parameter and $m_s$ a ``staggered''
mass-term, the sign of which alternates on the triangular sub-lattices of the hexagonal lattice. The mass term is added to remove zero-modes from the Hamiltonian and the limit $m_s \to 0$ is later taken.
The matrix $V$, which describes two-body interactions, must be positive-definite 
but otherwise may be chosen arbitrarily. We use the ``partially screened Coulomb potential'' in our simulations as described below.

A lattice path-integral representation for the grand-canonical partition function 
$Z=\Tr e^{-\beta H}$, in
which the operators $\ax,\axd,\bx,\bxd$ are replaced by Grassmann-valued 
field variables, can be derived by factorizing $e^{-\beta H}$ into $N_t$ terms
(implying a discretization error $\calo(\delta^2)$ where $\delta=\beta/N_t$)  and
inserting complete sets of of \emph{Fermionic coherent states}. 
The interaction term $\sim q_x V_{xy} q_y$ at first prevents the usual procedure of integrating
out Grassmann-fields via Gaussian integration, since it
contains fourth-powers of latter operators. 
These can be eliminated through \emph{Hubbard-Stratonovich}
transformation
\be
\exp\big\{ -
\frac{\delta}{2}\sum_{x,y}q_{x}V_{xy}q_{y} \big\}
\propto \int \big[ \prod_x \phi_x \big]\, \exp\big\{-\frac{\delta}{2}
\sum_{x,y} \phi_{x} V_{xy}^{-1} \phi_{y}
-i\,\delta
\sum_{x} \phi_{x} q_{x}\big\}~, \label{eq:hubbard1}
\ee
at the expense of introducing a dynamical scalar auxiliary field $\phi$ (``Hubbard field''),
which plays the role of a gauge field. In fact, $\phi$ can be understood as representing the scalar electric potential. A magnetic vector-potential does not appear since interactions
are taken to be instantaneous (which is a valid approximation since $v_F \ll c$).
The final result is the functional integral 
\be
Z =\int \mathcal{D}\phi
\, \det\big[ M(\phi) M^\dagger(\phi)  \big]
\exp \big\{-\frac{\delta}{2} \sum_{t=0}^{N_t-1} \sum_{x,y}
\phi_{x,t}V_{xy}^{-1} \phi_{y,t} \big\}~, \label{eq:partfunc1}
\ee
which is of a form that can be dealt with, using a standard Hybrid-Monte-Carlo 
algorithm to generate representative configurations of the Hubbard field $\phi$.
The Fermion-operator is given by
\be
M_{(x,t)(y,t')}~=~
\delta_{xy}(\delta_{tt'}-e^{-i\frac{\beta}{N_t}\phi_{x,t}} \delta_{t-1,t'})-
\kappa\frac{\beta}{N_t} \sum\limits_{\vec{n}} \delta_{y,x+\vec{n}}\delta_{t-1,t'}+ 
m_s\frac{\beta}{N_t} \delta_{xy} \delta_{t-1,t'}~.\label{eq:fermionmat1}
\ee
We choose to simulate rectangular graphene sheets with periodic boundary conditions. 
The dimensions $N_x$ and $N_y$ refer to the rhombic coordinate system
which spans the triangular sub-lattices.

\section{Semimetal-insulator transition}

A question which is of immediate significance to technological
applications (since electronic devices require a gate-voltage) is whether two-body interactions generate
a band-gap for some effective coupling constant $\alpha_{\textrm{eff}}$
which is smaller than the upper bound given by suspended graphene 
($\alpha_{\textrm{eff}}\approx 2.2$). This corresponds to electronic quasi-particles acquiring
a dynamical mass and at a microscopic level implies a spontaneous breaking of the symmetry
under exchange of the two triangular sub-lattices by formation of some condensate. Both
charge-density-wave (CDW) and spin-density-wave (SDW) formation are mechanisms which have been 
discussed in literature. As a result of studies which directly assumed or indirectly implied 
that the two-body interactions in graphene were essentially unmodified Coulomb-type
interactions (see e.g. Refs. \cite{Drut:2008rgArmour:2009vj,Buividovich:2012nx} and references
therein) it was predicted that the transition to a gapped phase happens for $\alpha_{\textrm{eff}} \gtrsim 1.0$,
well within the physically accessible region. This contradicted experiments 
which found graphene in vacuum to be a conductor \cite{Elias:2011xvMayorov}.

In Ref. \cite{Ulybyshev:2013swa} a Monte-Carlo study of the formation of an
anti-ferromagnetic condensate was conducted which suggested
screening of interactions by electrons in the lower $\sigma$-orbitals 
as a mechanism which moves the transition to $\alpha_{\textrm{eff}} \approx 3.14$ 
(to the unphysical region) and thus reconciles theory with experiment. Therein
explicit values for the on-site ($V_{00}$), nearest-neighbor ($V_{01}$),
next-nearest-neighbor ($V_{02}$) and third-nearest-neighbor ($V_{03}$) potentials from
calculations within a constrained random phase approximation (cRPA)
conducted in Ref. \cite{Wehling:2011df} were used. At long distances, it
was assumed that the potential falls off as $\sim 1/(\epsilon_\sigma \, r)$, where the constant
$\epsilon_\sigma \approx 1.41 $ was adjusted to match the $V_{03}$ term. A single lattice-size
($N_x=N_y=18, N_t=20$) and temperature ($\beta=2  \mathrm{eV}^{-1}$) were considered.

\begin{figure}
\begin{center}
\includegraphics[width=0.48\linewidth]{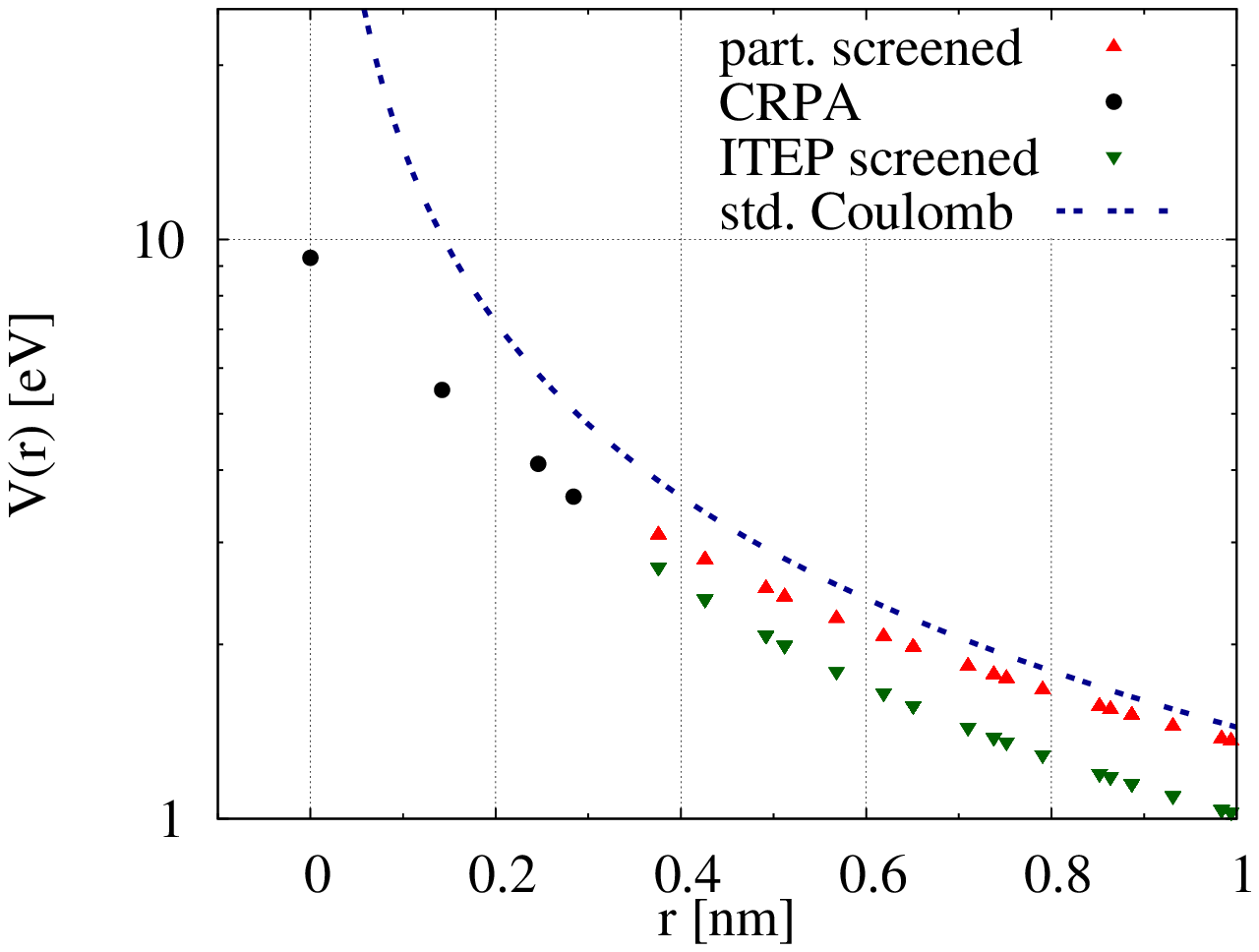}
\includegraphics[width=0.48\linewidth]{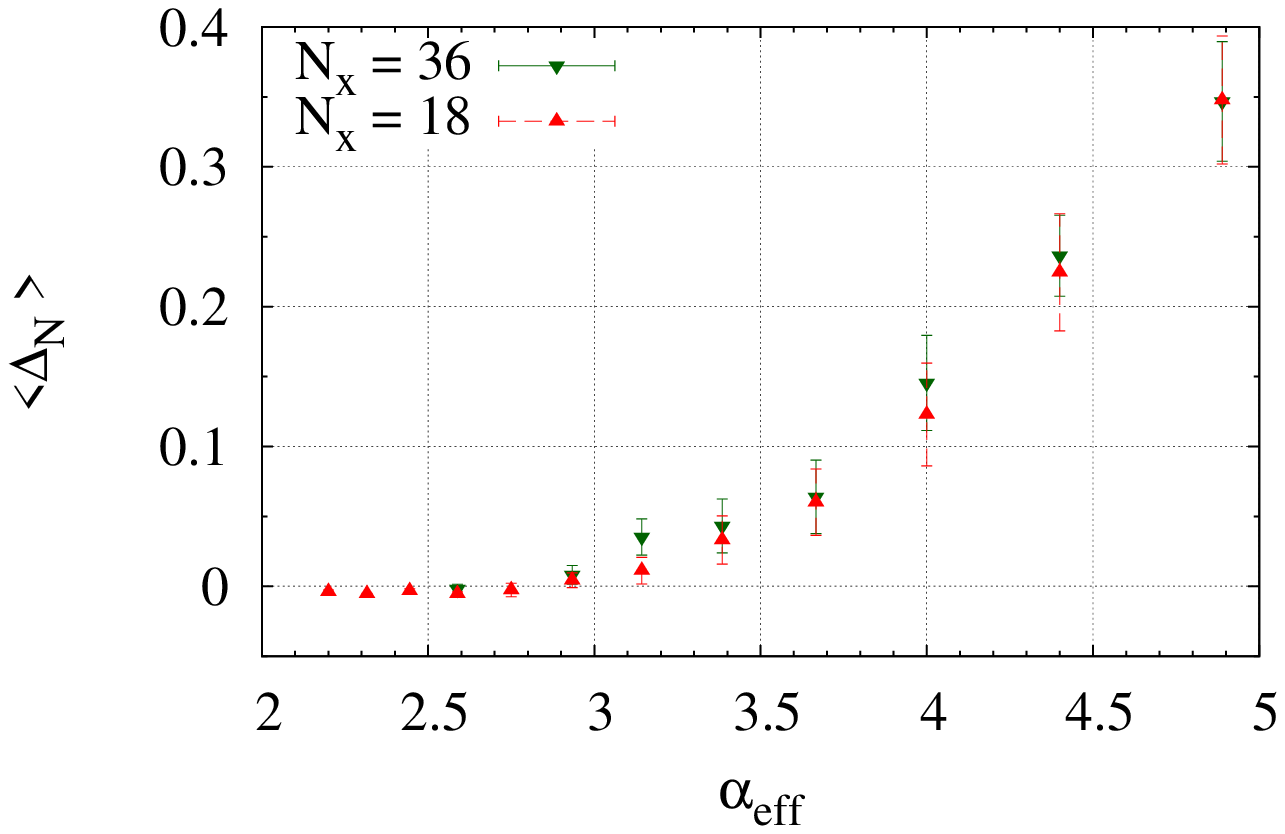}
\caption{\label{fig:partscreen1}
Left: Comparison of standard Coulomb potential, partially screened potential
and potential used in Ref. \cite{Ulybyshev:2013swa} (``ITEP screened''). 
Right: Order-parameter for gap-transition in the limit $m_s \to 0$.}
\end{center}
\end{figure}

Our work (Ref. \cite{Smith:2014tha}) aimed at eliminating residual doubts concerning
the long range tail. At long-distances we thus instead used a phenomenological model 
\begin{align}
\epsilon^{-1}_\sigma(\vec{k})=\frac{1}{\epsilon_1} 
\frac{\epsilon_1+1+(\epsilon_1-1) e^{-kd}}{\epsilon_1+1-(\epsilon_1-1) e^{-kd}}~,
\label{eq:dielectric1}
\end{align}
proposed in  Ref. \cite{Wehling:2011df} that describes a thin film of thickness 
$d=2.8 \; \textrm{\AA}$ ($\approx 1.41\cdot 10^{-3}\, \mathrm{eV}^{-1}$) with a dielectric
screening constant $\epsilon_1=2.4$ and computed a \emph{partially screened Coulomb potential}
via Fourier back-transformation
\begin{align}
V(\vec{r})&=\frac{1}{(2\pi)^2}\int_{\mathbb{K}^2} d^2k\,
\widetilde{V}_0(\vec{k})\, \epsilon^{-1}_\sigma (\vec{k})
\, e^{-i\vec{k} \vec{r}}~,\quad\quad \widetilde{V}_0(\vec{k})=(2\pi e^2)/k~,
\label{eq:screenedV1}
\end{align}
which smoothly goes over to an 
unscreened potential as $r \to \infty$ (see Fig. \ref{fig:partscreen1}, left). Simulations
using this potential were carried out, and the $\alpha_{\textrm{eff}}$ dependence of
the anti-ferromagnetic condensate
\be
\Delta_N =\frac{1}{N_x N_y}\big\{\sum\limits_{x\in X_A}( a^\dagger_{x} a_{x} + b^\dagger_{x} b_{x} )
-\sum\limits_{x \in X_B}( a^\dagger_{x} a_{x} + b^\dagger_{x} b_{x} )\big\},\label{eq:condens1}
\ee
was investigated for $N_x=N_y=18, N_t=20$ and $\beta=2 \mathrm{eV}^{-1}$ (the sums in Eq. (\ref{eq:condens1}) run over
coordinates in the two sub-lattices).
We found no substantial difference to the results of Ref. \cite{Ulybyshev:2013swa} and thus
concluded that the gap-transition is insensitive to the long-range tail of the
potential for this lattice-size. 

Fig. \ref{fig:partscreen1}, right, shows recent results for $\Delta_N(\alpha_{\textrm{eff}})$ which were obtained
on several hundreds of independent configurations of a $N_x=N_y=36,N_t=20$ lattice (for a range of $m_s$ which allowed
the limit $m_s\to 0$ to be taken) and compares them with
our published results obtained from $N_x=N_y=18$. A slight systematic effect is visible
which may indicate a small shift of the transition to a smaller value of $\alpha_{\textrm{eff}}$.
We take this result as evidence that finite-volume effects may still be present on
$N_x=N_y=18$ and that an infinite volume extrapolation is thus in order. It is conceivable
that the precise form of the long-range tail becomes more relevant for larger system-size. 

\section{Neck-disrupting Lifshitz transition}

\begin{figure}
\begin{center}
\includegraphics[width=0.97\linewidth]{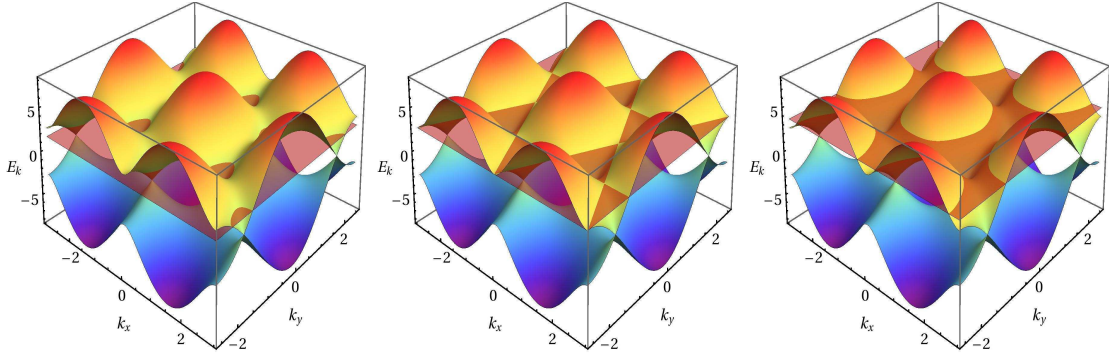}
\caption{\label{fig:Lifshitz1}
Band structure of tight-binding model, with Fermi-energy below
(left), exactly crossing (middle) and above (right) the $M$-points
(saddle points). Topology of the intersecting lines (``Fermi surface'') are different
in each case.}
\end{center}
\end{figure}

In the non-interacting limit the band structure of the tight-binding
Hamiltonian can be computed exactly \cite{Gusynin:2007ix}. It is known that the
valence and conduction bands possess within the first Brillouin zone,
(following the convention to  count points which are shared by cells fractionally)
three saddle-points, the so-called $M$-points. These points separate
the low-energy region, where the dispersion relation is approximately
linear, from a region in which electronic excitations are described
by the non-relativistic Schroedinger equation. By introducing a chemical potential
($H \to H + \mu \sum_{x,s} n_{x,s}$)
it is possible to shift the Fermi-energy across these points, which 
leads to a change of the topology of iso-energy lines
(see Fig. \ref{fig:Lifshitz1}). This is known
as a \emph{Neck-disrupting Lifshitz transition} and is accompanied
by a logarithmic divergence of the density of states \cite{Dietz:2013sga}
(``Van-Hove singularity'').

To date, little is known about the effect of two-body interactions on
the Lifshitz transition. 
Our goal is to clarify this issue through Monte-Carlo
simulation. In particular, it will be interesting to see how a small interaction
term affects the logarithmic scaling behavior. 
Adding a chemical potential leads to a Fermion sign-problem
however, since the Fermion operators are modified as
\be
M_{(x,t)(y,t')} \to M_{(x,t)(y,t')}
 + \mu \frac{\beta}{N_t} \delta_{xy} \delta_{t-1,t'}~,\quad\quad
M^\dagger_{(x,t)(y,t')} \to M^\dagger_{(x,t)(y,t')}
 - \mu \frac{\beta}{N_t} \delta_{xy} \delta_{t+1,t'}~, \label{eq:signprob1}
\ee
which implies that the phases in Eq. (\ref{eq:partfunc1}) no
longer cancel. As a first step it is reasonable to consider an
\emph{isospin chemical potential} 
$\mu_S\equiv s \mu$ (the sign of $\mu_S$ depends
on the direction of electron spin).
In the non-interacting limit it is clear that the Lifshitz
transition is blind to the sign of the spin.
For $\mu_S$ there is no sign-problem since positive terms are added
to both $M$ and $M^\dagger$.

We have so-far obtained preliminary results concerning the \emph{particle-number-susceptibility} associated with $\mu_S$ 
\be
\chi(\mu_S)  = \frac{1}{V \beta} \frac{d^2 \ln{Z}}{d\mu_S^2 }~,
\ee
which is related to the
density of states $\rho(\mu)$ (for $T=0$ they are exactly proportional)
and displays the characteristic
scaling behavior of the Lifshitz transition.
Hereby we restricted ourselves to the the non-interacting limit, 
where $\chi(\mu_S)$ can be computed directly from inversions of
the Fermion matrix on Gaussian noise vectors and no HMC updates are
necessary (since the Hubbard field is taken to zero). 
Furthermore, $\chi(\mu_S)$ can in fact be computed
exactly in this limit (by numerical
integration) from the retarded particle-hole polarization
(Lindhard) function (see Ref. \cite{Dietz:2013sga}).
For a given temperature $T$ it is (in infinite volume) given
by
\be
\chi(\mu_S)  =  \frac{1}{2T} \int_{BZ} \frac{d^2k}{4\pi^2} \Big[ \text{sech}^{2}\Big( \frac{\mu_S -E(\vec{k})}{2T} \Big) + \text{sech}^{2}\Big( \frac{\mu_S +E(\vec{k})}{2T} \Big) \Big]~, \label{eq:lindhard1}
\ee
where $E(\vec{k})$ is the dispersion-relation of the tight-binding theory and
the integral runs over the first Brillouin zone.
These results are important, since they
allow us to compare lattice results to direct
calculations and serve as a validation of our method. Simulations
at non-zero coupling are currently in progress.

\begin{figure}
\begin{center}
\includegraphics[width=0.98\linewidth]{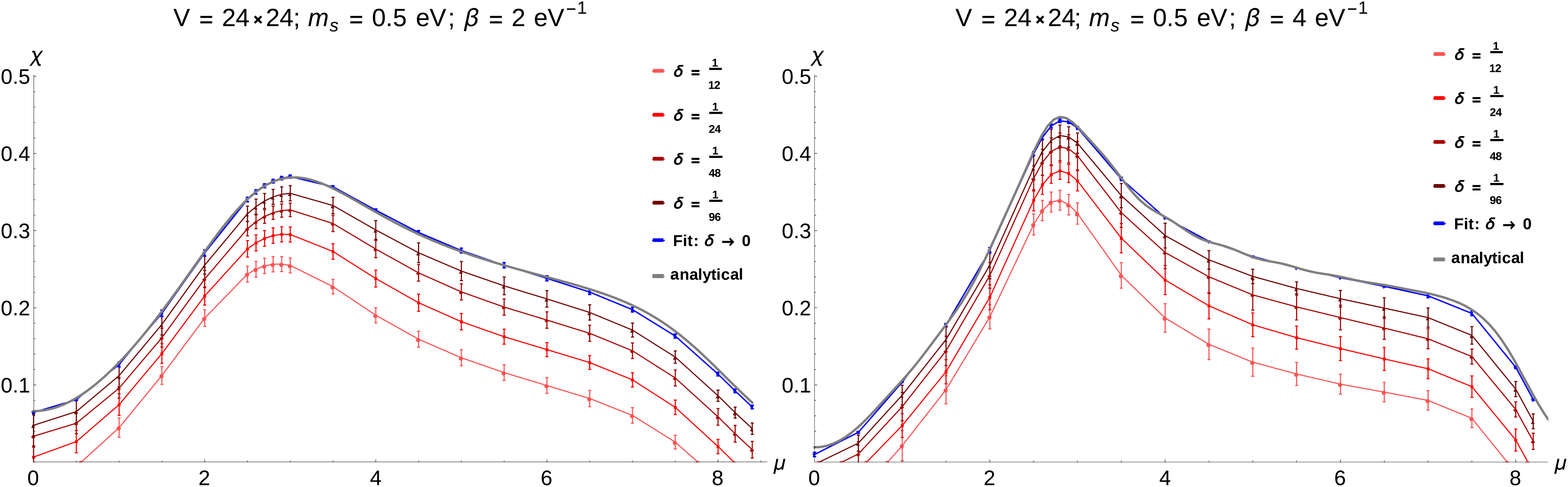}
\caption{\label{fig:Lifshitz2}
$\chi(\mu_S)$ in the non-interacting limit for $N_x=N_y=24$ and $m_s=0.5 \mathrm{eV}$. 
Results are obtained with different discretizations and the
limit $\delta \to 0$ is extrapolated. 
Two temperatures ($\beta=2;\,4\,\mathrm{eV}^{-1}$) are displayed. 
The solid line represents the exact results, obtained from
Eq. (\protect\ref{eq:lindhard1}). }
\end{center}
\end{figure}

Fig. \ref{fig:Lifshitz2} shows the susceptibility for two different temperatures
($\beta=2;\,4\,\mathrm{eV}^{-1}$) obtained from a  $N_x=N_y=24$ lattice with $m_s=0.5 \mathrm{eV}$.
We obtained data for different lattice-spacings and extrapolated to the
limit $\delta \to 0$. The results are compared to solutions of Eq. (\ref{eq:lindhard1})
(for finite volumes the integral is replaced by a sum over discrete momenta).
We find exact agreement within errors. Furthermore, we have confirmed that
the peak-height scales logarithmically with temperature as Eq. (\ref{eq:lindhard1}) also
predicts. 

\subsection*{Acknowledgments}

We have benefited from discussions with Pavel Buividovich, Maxim
Ulybyshev and David Scheffler. This work was supported by
the Deutsche Forschungsgemeinschaft within 
SFB 634, by the Helmholtz International Center for FAIR within the
LOEWE initiative of the State of Hesse, and the European Commission,
FP-7-PEOPLE-2009-RG, No. 249203.  All results were obtained using
Nvidia GTX and Tesla graphics cards.


\begin{thebibliography}{99}

\bibitem{CastroNeto:2009zzKotov} 
  A.~H.~Castro Neto et al.,
  Rev.\ Mod.\ Phys.\  {\bf 81}, 109 (2009);
  V.~N.~Kotov et al.,
  Rev.\ Mod.\ Phys.\  {\bf 84}, 1067 (2012)
  [arXiv:1012.3484].

\bibitem{Gusynin:2007ix}
  V.~P.~Gusynin et al.,
  Int.\ J.\ Mod.\ Phys.\ B {\bf 21} (2007) 4611
  [arXiv:0706.3016].

\bibitem{Drut:2008rgArmour:2009vj} 
  J.~E.~Drut and T.~A.~L\"ahde,
  Phys.\ Rev.\ Lett.\  {\bf 102}, 026802 (2009)
  [arXiv:0807.0834];
  Phys.\ Rev.\ B {\bf 79}, 165425 (2009)
  [arXiv:0901.0584].
  W.~Armour, S.~Hands and C.~Strouthos,
  Phys.\ Rev.\ B {\bf 81}, 125105 (2010)
  [arXiv:0910.5646].

\bibitem{Brower:2012zd} 
  R.~Brower, C.~Rebbi and D.~Schaich,
  PoS LATTICE 2011, 056 (2012)
  [arXiv:1204.5424];
  [arXiv:1101.5131].

\bibitem{Buividovich:2012nx} 
  P.~V.~Buividovich and M.~I.~Polikarpov,
  Phys.\ Rev.\ B 86, 245117 (2012)
  [arXiv:1206.0619].

\bibitem{Ulybyshev:2013swa} 
  M.~V.~Ulybyshev, P.~V.~Buividovich, M.~I.~Katsnelson and M.~I.~Polikarpov,
  Phys.\  Rev.\  Lett.\  111, 056801 (2013)
  [arXiv:1304.3660].

\bibitem{Smith:2013pxa} 
  D.~Smith and L.~von Smekal,
  PoS (LATTICE 2013) 048 [arXiv:1311.1130].

\bibitem{Smith:2014tha} 
  D.~Smith and L.~von Smekal,
  Phys.\ Rev.\ B {\bf 89}, 195429 (2014)
  [arXiv:1403.3620 [hep-lat]].

\bibitem{Dietz:2013sga}
  B.~Dietz, F.~Iachello, M.~Miski-Oglu, N.~Pietralla, A.~Richter,
  L.~von Smekal, and J.~Wambach,
  Phys.\ Rev.\ B\ {\bf 88}, 104101 (2013) 
  [arXiv:1304.4764].

\bibitem{Elias:2011xvMayorov}
  D.~C.~Elias et al.,
  Nature\ Phys.\ {\bf 7}, 701 (2011)
  [arXiv:1104.1396];
  A.~S.~ Mayorov et al.,
  Nano\ Lett.\ {\bf 12}, 4629 (2012),
  [arXiv:1206.3848].

\bibitem{Wehling:2011df} 
  T. O. Wehling et. al.,  
  Phys.\  Rev.\  Lett.\  106, 236805 (2011)
  [arXiv:1101.4007].



\end{thebibliography}
\end{document}